\documentclass[twocolumn,showpacs,preprintnumbers]{revtex4}
\usepackage{amssymb}
\usepackage{epsfig}
\usepackage{graphicx}
\usepackage{dcolumn}
\usepackage{bm}

\begin{document}

\title{Inhomogeneous exclusion processes with extended objects: \\The effect of defect locations}
\author{J. J. Dong, B. Schmittmann and R. K. P. Zia}
\affiliation{Center for Stochastic Processes in Science and Engineering, \\
Department of Physics, Virginia Tech, Blacksburg, VA 24061-0435, USA}
\email{jjdong@vt.edu}
\date{August 11, 2007}

\begin{abstract}
We study the effects of local inhomogeneities, i.e., slow sites of hopping rate $q<1$, in a totally asymmetric simple exclusion process (TASEP) for particles of size $\ell \geq 1$ (in units of the lattice spacing). We compare the simulation results of $\ell =1$ and $\ell >1$ and notice that the existence of local defects has qualitatively similar effects on the steady state. We focus on the stationary current as well as the density profiles. If there is only a single slow site in the system, we observe a significant dependence of the current on the \emph{location} of the slow site for both $\ell =1$ and $\ell >1$ cases. When two slow sites are introduced, more intriguing phenomena emerge, e.g., dramatic decreases in the current when the two are close together. In addition, we study the asymptotic behavior when 
$q\rightarrow 0$. We also explore the associated density profiles and
compare our findings to an earlier study using a simple mean-field theory. We then outline the biological significance of these effects.
\end{abstract}


\maketitle

\section{ \label{sec:Intro}{Introduction\protect\newline}}

A better understanding of non-equilibrium steady states in interacting
complex systems forms a critical goal of much current research in
statistical physics. In this pursuit, the totally asymmetric simple
exclusion process (TASEP) \cite{Krug,Derrida92,DEHP,S1993,Derrida,Schutz}
has played a paradigmatic role. It provides a nontrivial, yet exactly
solvable, example of phase transitions far from equilibrium, taking place even in one-dimensional (1D) lattices. At the same time, it also serves as the starting point for the modeling of many physical (driven diffusive) processes, such as translation in protein synthesis \cite{MG,LBSZia,TomChou}, inhomogeneous growth processes 
(e.g.~Kardar-Parisi-Zhang growth) \cite{KPZ,WolfTang} 
and vehicular traffic \cite{Chowdhury,Popkov}.

In its simplest version, the TASEP involves a single species of particles
hopping to nearest-neighbor sites, in one direction only, along a
homogeneous 1D lattice. Provided the destination site is empty, the rate for
the particle hop is fixed at $\gamma $ (typically chosen as unity without
loss of generality). With periodic boundary conditions, the steady-state
distribution is trivial \cite{Spitzer} but the full dynamics is quite
complex \cite{Gwa,Kim,BM}. With open boundary conditions, particles are
injected with rate $\alpha $ (in units of $\gamma $) at one end and drained
with rate $\beta $ at the other end. The competition of injection, transport
and drainage induces a nontrivial phase diagram in the $\alpha $-$\beta $
plane \cite{Krug,Derrida92,DEHP,S1993,Derrida,Schutz}, reflecting a highly
nontrivial steady state. Three phases are present: a maximum-current phase
for $\alpha ,\beta >1/2$, and a low- (high-) density phase for $\alpha
<\beta $, $\alpha <1/2$ ($\beta <\alpha $, $\beta <1/2$). Not surprisingly,
there are also rich dynamical aspects \cite{deGier,AZS}.

To model protein synthesis, each site on the lattice represents a codon on the messenger RNA (mRNA), and the particles represent the ribosomes.
Injection, hopping, and drainage are associated respectively with
initiation, elongation, and termination in biological terms. The quantity of interest, namely, the (steady-state) protein production rate, is identical to the (stationary) particle current. Clearly, the simple TASEP falls short of the biological system in several significant aspects. One is that an individual ribosome ``covers'' several codons \cite{MG,Heinrich,Kang}, as opposed to a particle occupying only a single site. Another is that, in all naturally occurring mRNAs, the codons carry genetic information and therefore
necessarily form an \emph{inhomogeneous} sequence. Thus, the elongation rate of a ribosome is unlikely to be uniform; instead, the hopping rate, $\gamma_{i}$, of a particle becomes a function of the site $i$. For example, it is well known that translation slows down at specific codons (see, e.g. \cite{Solomovici,Stenstrom,TomChou,Chou9, Chou10}), with potentially significant consequences for protein production rates. Indeed, the steady-state current may depend sensitively on not only the \emph{frequency} of each codon's occurrence, but also the \emph{order} of their appearance in the sequence. Both of these issues -- extended objects and inhomogeneous rates -- have been addressed recently in separate contexts which we summarize briefly in the following.

The results associated with inhomogeneous (quenched random) rates fall into two broad categories, in the sense that the randomness can be associated with the particles 
\cite{EnaudDerrida16,EnaudDerrida23,EnaudDerrida24} or
with the sites. Randomness of the former type is more relevant for vehicular traffic where it accounts for a variety of driver preferences. In contrast, the disorder in the protein case is clearly \emph{site-dependent}, leading to \emph{spatially} non-uniform hopping rates $\gamma _i$. Restricting ourselves to this class, we can consider the effect of having a whole \emph{distribution}, or very \emph{specific} configurations, of $\gamma _i$. Starting from given distributions, two groups \cite{TripathyBarma,Harris}
studied the resulting disorder-average in periodic systems. To mention just one significant effect, the current-density diagram develops a plateau: limited by the smallest rate in the system, the current becomes independent of density over a range of densities. Harris and Stinchcombe \cite{Harris} also extended this work to open systems. While these studies may be of some interest to \emph{mixtures} of many different mRNAs, our primary interest here is to understand how the production rate of a specific protein is associated with a specific genetic sequence. As a first step towards a solution, we adopt the approach of several other studies \cite{Kolo,HadN,LBS,TomChou}, by focusing on the effects of a few \emph{localized} inhomogeneities, i.e., hopping rates which are uniform \emph{except} at a handful of sites \cite{note1}. As a synthesis of these studies, we will
explore in some detail the consequences of having extended objects and
locating one or two slow sites at a variety of positions on the lattice. In this manner, by introducing more and more sites with a range of rates, we hope to understand inhomogeneities in a systematic way, setting the stage for further investigations of the translation process.

A full comprehension of the effects of slow sites on the particle current may have potentially significant applications in biotechnology. While there are 64 distinct codons, proteins are chains composed of just 20 amino acids. So, many different mRNAs (codon sequences) can code for the same particular protein. Moreover, the amino acid is incorporated into the growing chain by an important intermediary, the so-called transfer RNA (tRNA), which carries the complementary anticodon. It turns out that the mapping between codons and tRNAs is also not precisely 1-1. For example, in \textit{E.~coli}, the genetic code actually involves 61 sense codons and about 46 tRNAs with associated anticodons \cite{Neidhardt}. Meanwhile, for a given mRNA
sequence, the protein production rate is often modeled in terms of
(generally accepted) charged-tRNA (aminoacyl-tRNA, or aa-tRNA)
concentrations \cite{Solomovici}, so that different sequences can result in \emph{different} production rates for the \emph{same} protein. By elucidating how the spatial distribution of defects, especially of bottlenecks, affects translation rates, we can pinpoint those clusters of codons which are likely to have the most significant effect on the production rate of the associated protein. Exploiting the degeneracy in the mapping from mRNA sequence to protein, we can provide guidance as to how a few selected, local modifications of the mRNA can optimize the production rate of a given protein.

Our paper is organized as follows. In Section \ref{sec:model}, we define the model and provide a more detailed description of previous work, concerning exclusion processes with extended objects or spatially inhomogeneous rates. A brief discussion of a previous mean-field analysis is also included. In Section \ref{sec:MC}, we present our Monte Carlo results. We focus especially on the implications of having extended objects by varying the particle size. We first consider the interaction between one slow site and the system boundary, and, motivated by the resulting findings, turn to the interactions between \emph{two} slow sites. This provides new insights for genes containing clusters of slow codons, which occur frequently in, e.g., 
\textit{E.~coli, Drosophila}, yeast and primates 
\cite{Chou11,Chou14,TomChou}. In Section \ref{sec:MF}, a complete investigation of systems with inhomogeneities is presented using a mean-field approach. Section \ref{sec:SUM} contains our conclusions and a summary of open questions.

\section{\label{sec:model}{Model specifications and known results}}

The TASEP is defined on a 1D lattice of $N$ sites. We introduce an index $i=1,2,...,N$ to label the sites. Each site (codon) is either occupied by a single particle (ribosome) of length $\ell $ (in units of sites) or empty. A microscopic configuration of the system can be uniquely characterized in terms of a set of occupation variables, $\left\{ n_i\right\} $, taking the value 1(0) if site $i$ is occupied (empty). Of course, the extended nature of the particle induces strong correlations in $\left\{ n_i\right\} $, in the sense that a single ribosome always covers $\ell $ \emph{consecutive} sites. Yet, at any given time, only one of the covered codons is being ``read'' (i.e., the codon is ``covered by the aminoacyl site'', or A site, of the ribosome) and translated into an amino acid. Here, we refer to the associated location on the ribosome as the ``reader'' (of the genetic code).
For our purposes, it is not essential which one of the $\ell $ sites is labeled as the reader, and so we follow the convention in \cite{LBSZia} and choose the first (leftmost) site. Hence, the statement ``a ribosome (or particle) is located at site $i$'' implies that the reader is located at site $i$ and the subsequent 
\begin{equation}
\bar{\ell}\equiv \ell -1  \label{ell-bar}
\end{equation}
sites are also occupied. Naturally, the position of the reader determines the elongation rate, i.e., $\gamma _i$, since the ribosome must wait for the arrival of the aa-tRNA with the $i$-specific anticodon before it can move to the next site. Clearly, the reader locations can also be used to label a microscopic configuration, i.e., we can define the reader occupation number at site $i$ as $r_i$. The sets $\left\{ n_i\right\} $ and $\left\{ r_i\right\} $ are uniquely related to each other. Moreover, due to the extended size of a particle, strict \emph{constraints} are built in 
(e.g., $r_i=1$ implies $r_{i+1}=...=r_{i+\ell -1}=0$ 
and $n_i=...=n_{i+\ell -1}=1$). As a consequence, neither set can be arbitrary and serious correlations arise as soon as $\ell >1$ 
\cite{note2}.

\begin{figure}[tbp]
\includegraphics[height=1in,width=3.5in]{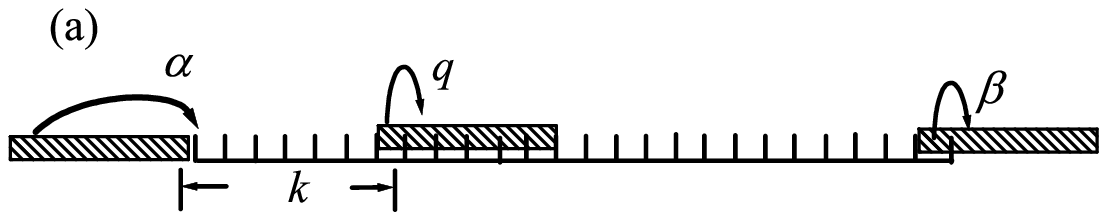} 
\includegraphics[height=1in,width=3.5in]{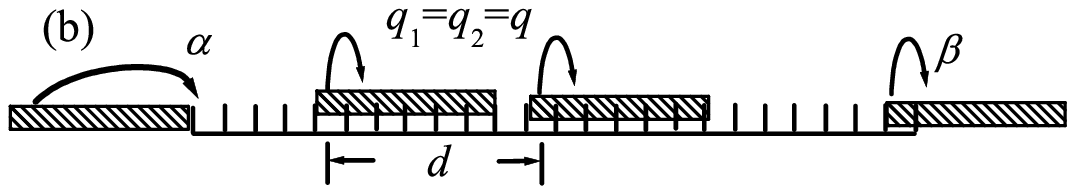}
\caption{Sketch of a TASEP for particle size $\ell =6$ with (a) a single slow site at position $k$, with rate $q$, and (b) two slow sites with rate $q $, separated by a distance $d$.}
\label{fig:one-slow}
\end{figure}

In our simulations, we adopt a random sequential updating scheme and keep a list of locations of readers. In addition, the site $i=0$ is always occupied by a ``virtual reader,'' which accounts for particles entering the system (initiation). At the beginning of each Monte Carlo step (MCS), we first find the number of particles in the system and label it $M$. Then, we randomly select an entry from this list of 
$M+1$ readers. If the chosen reader is virtual (i.e., $i=0$), a new particle enters the lattice with probability $\alpha $, \emph{provided} all the first $\ell $ sites are empty. If the chosen reader is real, say, at site $i>0$, the associated particle is then moved to site $i+1$ with probability $\gamma _i$, \emph{provided} site $i+\ell $ is empty. With this notation, we can also write the initiation and termination probabilities ($\alpha $ and $\beta $) as 
$\gamma _0$ and $\gamma _N$, respectively. To be complete, the sites beyond the lattice are by definition empty, so that once a particle reaches $N-\ell +1$, it will not experience steric hindrance 
(see Fig.~\ref{fig:one-slow} for a sketch of this process). These processes have been termed ``complete entry'' and ``incremental exit'' \cite{Lakatos}. Other entry and exit rules can be considered, but are believed to be inconsequential provided $\ell /N\ll 1$. Each MCS consists of $M+1$ such attempts, giving an even chance, on average, for each particle (ribosome) in the system to elongate or terminate, as well as for an initiation event to occur.

Starting with an empty lattice, we typically discard $2\times 10^{6}$ MCS to ensure that the system has reached the steady state. Unless otherwise noted, good statistics result if we average over least $2\times 10^{4}$ measurements, separated by $100$ MCS in order to avoid temporal correlations. Such steady state averages will be denoted by $\left\langle ...\right\rangle $. To reduce the number of parameters in the model, we study systems with $\alpha =\beta =\gamma _{i}=1$, 
\emph{except} at one or two sites. The system sizes ($N$) range from $200$ to $1000$, with most data taken from $N=1000$.

To characterize the state of the system, we monitor several observables. The most obvious is $\rho _i^{\text{r}}\equiv \left\langle r_i\right\rangle $, a quantity we will refer to as the ribosome (or reader, or particle) density. Of course, $\sum_i\rho _i^{\text{r}}$ is just the average number of particles in the system (i.e., ribosomes on the mRNA). Thus, the overall particle density 
$N^{-1} \sum_i\rho _i^{\text{r}}$ is bounded above by $1/\ell $. Another interesting variable 
$\rho _i\equiv \left\langle n_i\right\rangle $, labeled as the ``coverage density'', is the probability that site $i$ is covered by a particle (regardless of the location of the reader). Needless to say, the profile for the vacancies is given by the local hole density, 
$\rho _i^{\text{h}} \equiv 1-\rho _i$. The overall coverage
density, $ N^{-1} \sum_i\rho _i$, may reach unity and provides a good
indication of how packed the system is. The two profiles are related by 
\begin{equation}
\rho _i=\sum_{k=0}^{\ell -1}\rho _{i-k}^{\text{r}}
\end{equation}
{with} the understanding $\rho _i^{\text{r}}\equiv 0$ for $i\leq 0$.

A quantity of great importance to a biological system is the 
steady-state level of a given protein. If we assume that the degradation rates are (approximately) constant, i.e., independent of protein concentration, then these levels are directly related to the protein production rates. In our model, such a rate is just the average particle current $J$, defined as the average number of particles exiting the system per unit time. In the steady state, it is also the current measured across any section of the lattice. For simplicity and to ensure the best statistics, we count the total number of particles which enter the lattice over the entire measurement period (at least $2\times 10^6$ MCS in most cases).

In this study, we focus on two simple types of inhomogeneities: one or two ``slow'' sites (Fig.~\ref{fig:one-slow}). Their locations specify the only inhomogeneities in the rates.

\textit{One slow site}, at position $k$. We denote $\gamma _k$ by $q$ ($<1$). This corresponds to a bottleneck in the lattice. We are especially interested in the dependence of the current, denoted by $J_q(k)$, on the parameters $q$ and $k$.

\textit{Two slow sites}, at positions $k_1$ and $k_2$ with separation 
$d\equiv (k_2-k_1)$ and rates $q_{1,2}\equiv \gamma _{k_{1,2}}.$ We find that, when $q_1\neq q_2$, the current is controlled mainly by the smaller of the two, with little dependence on $d$, in agreement with a simple mean-field theory to be discussed in Section \ref{sec:MF}. Therefore, most of our attention will be devoted to the case with 
$q_1=q_2\equiv q<1$.
Moreover, we choose to limit our study to both sites being far from the
boundaries. Then, the current is insensitive to their average position $(k_2+k_1)/2$, and we can investigate $J_q(d)$. Note that these are precisely the systems studied in \cite{TomChou}, except that we consider particles with a \emph{range} of sizes: $\ell =1,2,4,6,$ and $12$. While there are qualitative similarities, we will discuss the \emph{quantitative} differences due to $\ell >1$, as well as interesting phenomena associated with the density profiles.

\begin{figure}[tbp]
\includegraphics[height=3in,width=3in]{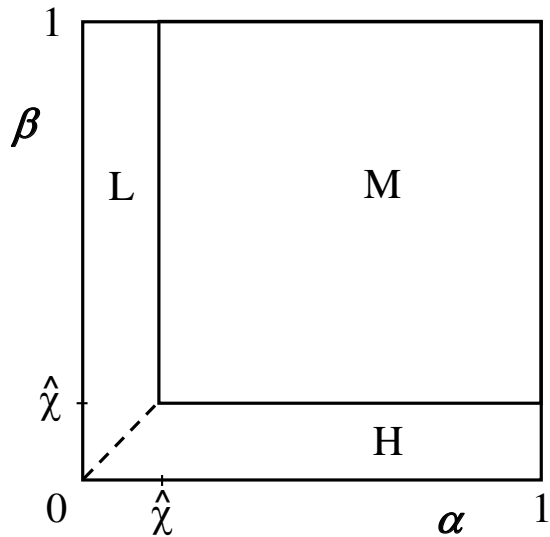}
\caption{Phase diagram for an ordinary TASEP. On the dashed line, the H and L phases coexist. }
\label{fig:phase-ext}
\end{figure}

Let us now provide the context of our work by briefly reviewing some related
earlier studies. The homogeneous case ($\gamma _1=...=\gamma _{N-1}=1$) with 
$\ell =1$ is exactly soluble \cite{Derrida92,DEHP,S1993,Derrida,Schutz}, and
displays three phases in the $\alpha $-$\beta $ phase diagram. For $\ell >1$%
, no exact solutions exist\cite{noExact}. Analytic approximations using
various mean-field approaches \cite{MG,Heinrich,Lakatos,LBSZia,SS} predict
the presence of the same phases, though the phase boundaries depend on $\ell 
$ (Fig.~\ref{fig:phase-ext}) through the combination \cite{LBSZia}
\begin{equation}
\hat{\chi}\equiv \frac 1{1+\sqrt{\ell }}  \label{chi-hat}
\end{equation}
Monte Carlo studies \cite{Lakatos,LBSZia} largely confirm these conclusions.

The three phases carry different currents and display distinct density
profiles \cite{Derrida92,DEHP,S1993,Derrida,Schutz,Lakatos,LBSZia,LBS}.
Apart from ``tails'' near the boundaries, the (coverage) density profiles approach uniform bulk values in the thermodynamic limit, i.e., $\rho_i\rightarrow $ $\rho _{bulk}$, for $1\ll i\ll N$. 
For $\alpha <\hat{\chi}$ and $\alpha <\beta $, the system is in a low-density phase (L), characterized by 
$\rho _{bulk}=\ell \alpha /\left( 1+\alpha \bar{\ell}\right) $ and 
$J=\alpha (1-\alpha )/\left( 1+\alpha \bar{\ell}\right) $. A
high-density phase (H) prevails for $\beta <\hat{\chi}$ and 
$\beta <\alpha $, with bulk density $\rho _{bulk}=1-\beta $ and current $J=\beta (1-\beta )/\left( 1+\beta \bar{\ell}\right) $. For 
$\alpha ,\beta >\hat{\chi}$, the system is in a maximum-current phase (M), where $\rho _{bulk}=1-\hat{\chi}$ and $J=\hat{\chi}^2$. On the $\alpha =\beta <\hat{\chi}$ line (dashed line in Fig.~\ref{fig:phase-ext}), the system consists of two macroscopic regions, characterized by a low (high) density region near the entry (exit) point. The two regions are joined by a shock front that performs a random walk. This is often referred to as the ``shock phase'' (S). Table \ref{tab:J-rho}
summarizes the $J - \rho _{bulk}$ relation for TASEP with extended objects. 
\begin{table}[tbp]
\caption{$J$-$\rho _{bulk}$ relation for particles of size $\ell $ 
($\bar{\ell}\equiv \ell -1$).}
\label{tab:J-rho}%
\begin{ruledtabular}
\begin{tabular}{ccc}
\mbox{phase}&\mbox{current $J$}&\mbox{bulk density $\rho_{bulk}$}\\
\hline\\
\vspace{0.2cm}
L & $\alpha(1-\alpha )$/ (1 + $\alpha \bar{\ell}$) & $\ell \alpha$ /(1 + $\alpha \bar{\ell}$)\\
\vspace{0.2cm}
H & $\beta (1-\beta )$/ (1 + $\beta \bar{\ell}$)  & $1-\beta $\\
\vspace{0.2cm}
M & $\hat{\chi}^2$ & $1-\hat{\chi}$\\
\end{tabular}
\end{ruledtabular}
\end{table}
There is good agreement between simulations (with $\ell \leq 12$) and
analytic results for these bulk quantities \cite{Lakatos,LBSZia}. The
details of the profile for $\ell >1$, especially near the lattice
boundaries, are less well understood. While periodic structures (of period $\ell $) can be expected, mean-field theories 
\cite{MG,Heinrich,Lakatos} were successful in capturing only a limited part of the phenomena observed. We will return to these considerations in Section \ref{sec:MF}.
Beyond homogeneous systems, several studies introduced one or more
``impurities'' into TASEP with periodic boundary condition. A single
``slow'' site induces a shock in the density profile with some interesting statistics \cite{Janowsky,JandL,HadN13,HadN14,HadN15}. Subsequently, generalizations to systems with a finite fraction of slow sites, randomly located, were also investigated \cite{TripathyBarma}. For the richer case of the open boundary TASEP 
\cite{Kolo,HadN,LBS,TomChou,DSZ}, Kolomeisky focused on point particles ($\ell =1$), with a \emph{single} impurity at the \emph{center} of the lattice \cite{Kolo}, so as to mimic a defect situated deep in an infinitely long system. The consequences of the defect having both faster ($q>1$) and slower ($q<1$) rates were explored. By matching two ordinary TASEPs across the defect, the properties of such systems in the $\alpha $-$\beta $ plane can be well described \cite{Kolo,DSZ}. While a fast site has no effect on the phase diagram, a slow site leads to a shift of the M-H and M-L phase boundaries to $q$-dependent, smaller values of $\alpha $ and $\beta $. The density profiles are quite sensitive to the existence of a defect site \cite{Kolo,TomChou}. Kolomeisky's approach was generalized to the $\ell =12$ case in \cite{LBS}, with similar levels of success. Below, we will provide further details of this work, on which we base much of the analysis of our problem. Ha and den Nijs also studied the $\ell =1$ open boundary TASEP with a single defect at the center \cite{HadN}. Focusing on the multicritical point $\alpha =\beta =1/2$, they were mainly interested in the so-called ``queueing transition'' and its critical properties. Detailed results of density profiles, such as power law behavior and critical exponents, were obtained in the region $q\cong q_c$. Here, $q_c$ denotes the critical value of $q$ below which the bulk density in front of the slow site deviates from the density behind the blockage. By contrast, our focus here is essentially that of \cite{TomChou,DSZ}, namely, how does the \emph{number} and the \emph{locations} or \emph{spacings} of the slow sites affect the current through the system? In the single slow site case, \cite{TomChou} investigated mainly the overall current as a function of $q$, while \cite{DSZ} analyzed the current as a function of $k$ (the distance between the defect and the entry point). For the case of two defects, both studies find that the spacing between them plays a significant role for the current. A
finite-segment mean-field theory in \cite{TomChou} provides excellent
agreement with data. In particular, clustered defects reduce the current much more effectively than well separated ones. In this sense, we can regard these investigations as exploring the ``interactions'' between the slow site(s) and/or the boundaries. Since both studies are restricted to point particles ($\ell =1$), our intent is to explore the effects of having extended objects (with $\ell \leq 12$). Though we expect \emph{qualitatively} similar behavior, as pointed out in \cite{TomChou}, we also find noteworthy \emph{quantitative} differences. Finally, we should mention that larger numbers of defects do not lead to significantly different effects \cite{TomChou}, so that we limit ourselves to one or two slow sites here.

\section{\label{sec:MC}Monte Carlo results}

In this section, we present our Monte Carlo results. For convenience, we use a consistent color coding scheme (online only) for the various particle sizes, as specified in Table \ref{tab:color}.
\begin{table}[tbh]
\caption{Color coding scheme}
\label{tab:color}
\begin{tabular}{|c|c|}
\hline
\mbox{size $\ell$} & \mbox{online color} \\ \hline
1 & black \\ 
2 & red \\ 
4 & brown \\ 
6 & green \\ 
12 & blue \\ \hline
\end{tabular}
\end{table}
The data here consist of the overall currents and the density profiles of both coverage and ribosomes. Our focus will be how these quantities depend on $q$, $k$ (for the case with a single defect), and $d$ (for the case with two defects). Although the profiles are difficult to extract experimentally, the reader profiles will be of interest in subsequent studies involving real gene sequences, since they provide information on how frequently a particular tRNA is bound to the mRNA. By contrast, the currents are easily measurable and our results here may generate more immediate interest.
\subsection{One slow site}

\begin{figure}[h]
\hspace{-0.5cm} %
\includegraphics[height=3in,width=3.5in]{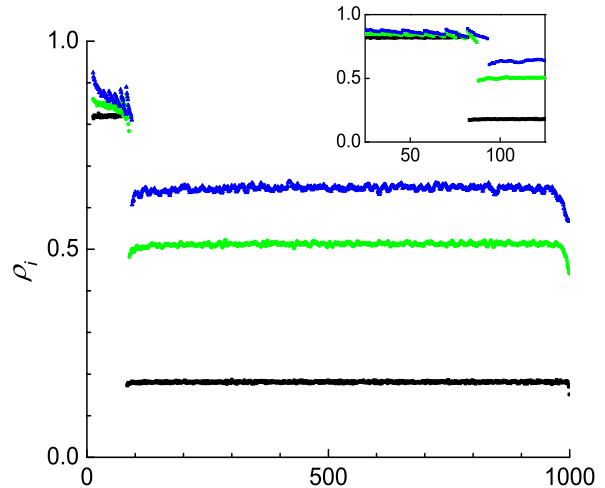} 
\vspace{
-0.6cm}
\caption{(Color online) Coverage density profiles with one slow site of $q=0.2$ at 
$k=82$. $\ell=1, 6, 12$ (from bottom to top in both subsections) and $N=1000$. The inset is a magnified view of the $i\in\left[1,150\right] $ interval, to expose the period $\ell $ structures.}
\label{fig:one-profile}
\end{figure}

We begin by placing one slow site on the lattice as in 
Fig.~\ref{fig:one-slow}(a). Fig.~\ref{fig:one-profile} shows several coverage density profiles for a typical choice of parameters: $N=1000$, $q=0.2$, and $k=82$ with $\ell =1,6,12$. As expected, we observe 
pile-ups of particles due to the blockage -- a high (low) density region before (after) the bottleneck in all three cases. However, due to the lack of ordinary particle-hole symmetry in the $\ell >1$ cases, the average densities on either side of the slow site are not symmetric around $0.5$. Instead, they are \emph{roughly} related through the $J$-$\rho _{bulk}$ relations in the H and L phases, summarized in Table \ref{tab:J-rho}. In detail, the profiles are quite different: The ``tails'', i.e., the deviations from the bulk values, are quite noticeable in the vicinity of both the slow site and the edges of the
system for the $\ell =6,12$ cases. The inset exposes more clearly that there are period $\ell $ structures in the profiles 
\cite{MG,Heinrich,Lakatos}, especially just before the slow site. 

\begin{figure}[tbp]
\hspace{-0.5cm} %
\includegraphics[height=3in,width=3.5in]{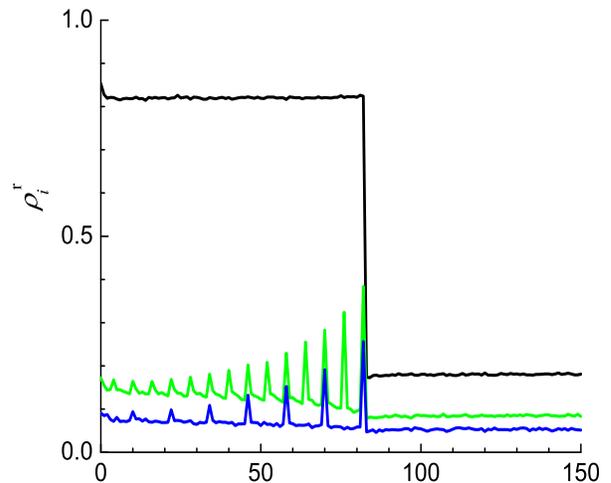} 
\vspace{-0.6cm}
\caption{(Color online) Ribosome density profiles with one slow site of $q=0.2$ at 
$k=82$. $\ell=1, 6, 12$ (from top to bottom in both subsections) and $N=1000$. Only the first 150 lattice sites are shown.}
\label{fig:one-ribo-profile}
\end{figure}

\begin{figure}[tbp]
\hspace{-0.5cm} \includegraphics[height=3in,width=3.5in]{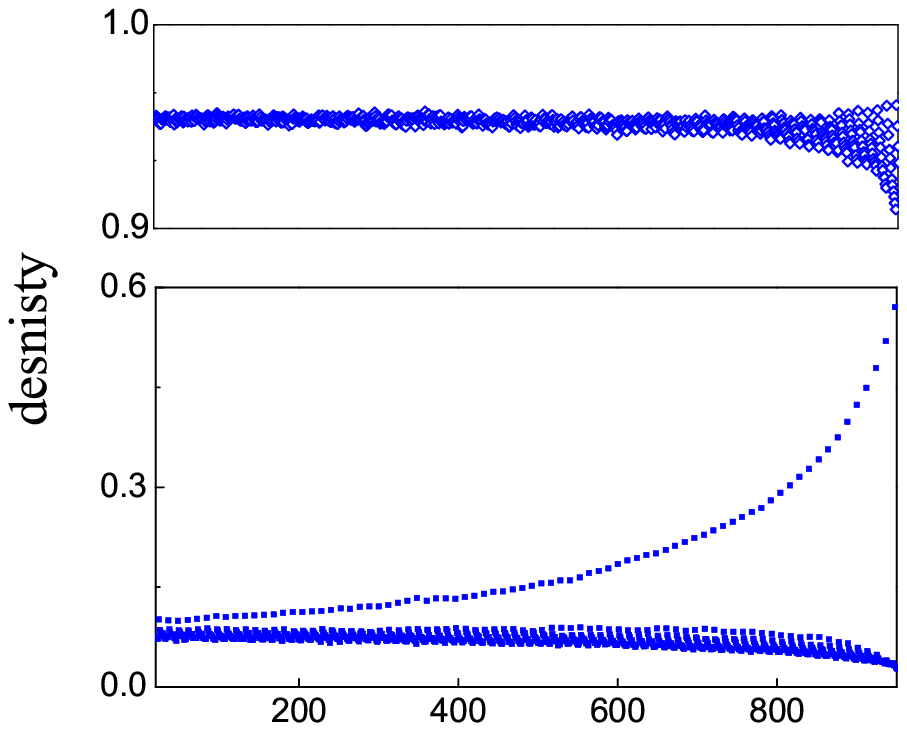} 
\vspace{-0.5cm}
\caption{(Color online) Coverage density profile (top) and ribosome density profile
(bottom) with one slow site of $q=0.05$ at $k=948$. $\ell=12$ and $N=1000$.}
\label{fig:one-profile-cf}
\end{figure}

A more dramatic difference between point particles and extended objects
emerges when we plot the ribosome density $\rho _i^{\text{r}}$, in 
Fig.~\ref{fig:one-ribo-profile}, corresponding to the inset in 
Fig.~\ref{fig:one-profile}. Similar to profiles in 
\cite{MG,Heinrich,Lakatos}, we find distinct period $\ell $ structures before the slow site. While the reader ``waits'' to pass the blockage, the readers of the following particles tend to catch up and pause at sites $k-n\ell $, where $n=1,2,..$. The ``tails'' are even more marked than those in Fig.~\ref{fig:one-profile}. To emphasize the difference between the reader and coverage profiles ($\rho_i^{\text{r}}$ and 
$\rho _i$), we show a case with $q=0.05,k=948,\ell =12$
in Fig.~\ref{fig:one-profile-cf}. Though both profiles contain the same
information, we see that $\rho _i^{\text{r}}$ (lower plot) is far more
sensitive than $\rho _i$ (upper plot) in showing the very long tails 
($\sim 1000$ in this example) hidden in the collective behavior of the particles. At present, the crucial ingredients that control the
characteristic decay length of the $\rho _i^{\text{r}}$-envelopes have not yet been identified. Certainly, these very large length scales are
completely absent from the $\ell =1$ systems deep within the H/L phases.


\begin{figure}[tbp]
\hspace{-0.5cm} \includegraphics[height=3in,width=3.5in]{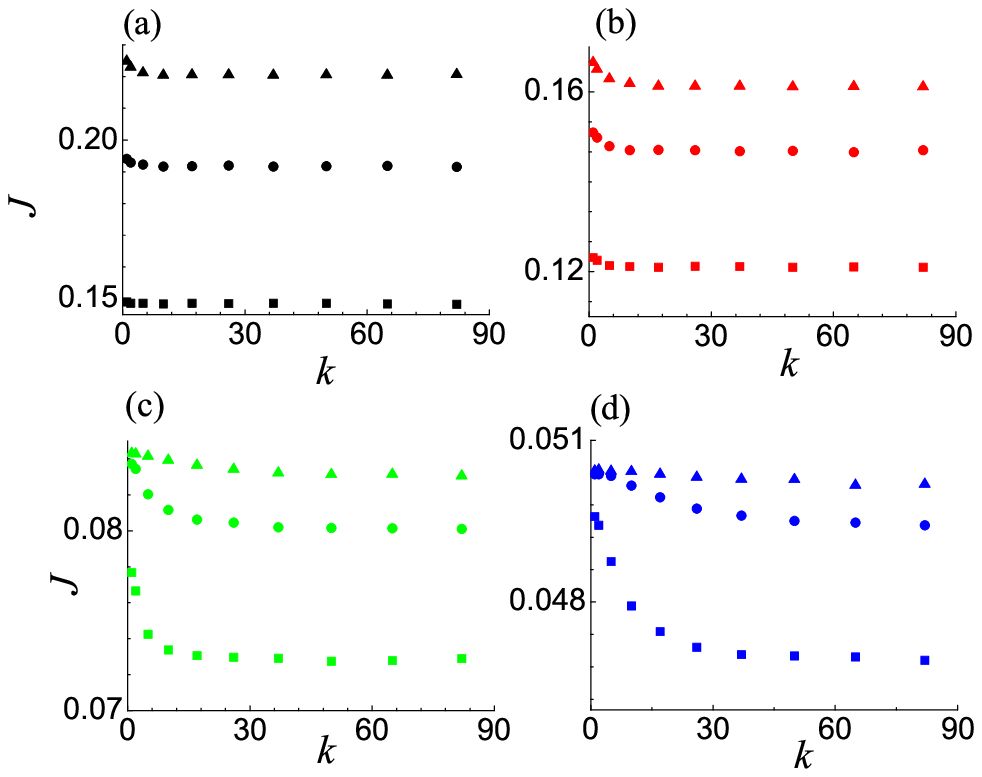} 
\vspace{-0.6cm}
\caption{(Color online) $J_q(k)$ as a function of the location $k$ of the slow site for $q=0.2$ (lower set of squares); $0.3$ (middle circles) and $0.4$ (upper triangles). (a) $\ell =1$; (b) $\ell =2$; (c) $\ell =6$; (d) $\ell =12$. In all cases, $N=1000$.}
\label{fig:one-J}
\end{figure}
\begin{figure}[tbp]
\hspace{-0.5cm} \includegraphics[height=3in,width=3.5in]{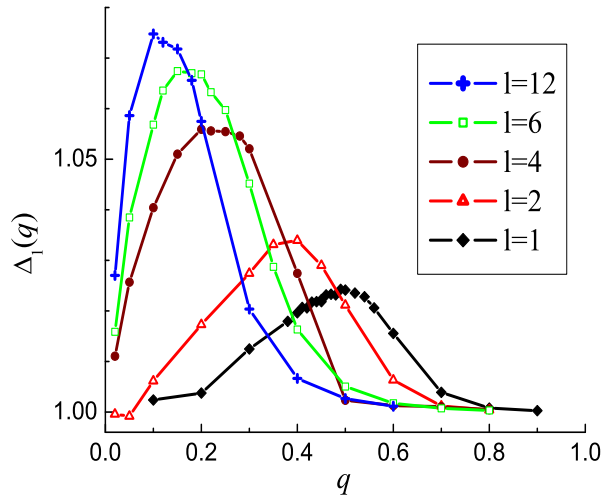} 
\vspace{-0.6cm}
\caption{(Color online) $\Delta _1(q)$ for $\ell =1,2,4,6$ and $12$.}
\label{fig:one-diff}
\end{figure}

As for the current, Fig.~\ref{fig:one-J} illustrates its dependence on $q$, $k$, and $\ell $. Not surprisingly, the current is limited by the bottleneck and therefore varies monotonically with $q$. It is also reduced if the particle size increases, an effect that can be traced mainly to the particle density being effectively lower by the factor $\ell $. For point particles, the current is not very sensitive to the location of the slow site. The enhancement as $k$ approaches the boundary of the system -- referred to as the ``edge effect''\cite{DSZ} -- is quite small. For larger $\ell $, the enhancement is much more pronounced, especially for smaller $q$. Whatever the magnitude, in all cases the current increases monotonically as the slow site is located closer and closer to the entry point. For $\ell =1$, particle-hole symmetry is manifest in the microscopic dynamics, so that the symmetry of $J_q(k)$ under $k\rightarrow N+1-k$ inversion, is obvious 
\cite{DSZ}. For $\ell >1$, the density profiles confirm the \emph{lack} of this particle-hole symmetry very clearly. Correspondingly, there is a systematic asymmetry in the current: $J_q(k)=J_q(N+1-k)$ is satisfied only for $k\lesssim \ell $. The origin of this behavior is not well understood.

The edge effect, and specifically its dependence on $q$ and $\ell $, can be quantified by the ratio: 
\begin{equation}
\Delta _1(q)=\frac{J_q(k=1)}{J_q(k\rightarrow \infty )}\,\,
\label{delta-J-1}
\end{equation}
Fig.~\ref{fig:one-diff} shows that $\Delta _1(q)$ depends on $q$ in a
nontrivial way. The maxima of $\Delta _1(q)$ occur at lower values of $q$ as $\ell $ increases, reminiscent of the behavior of the phase boundary between M and L/H. With appropriate scaling, the curves of $\Delta _1(q)$ can be collapsed for large $\ell $'s. From the biological perspective, the edge effect is not easily observable since the current enhancement is less than $10\%$ for the relevant $\ell $.

\begin{figure}[tbp]
\hspace{-0.5cm} \includegraphics[height=3in,width=3.5in]{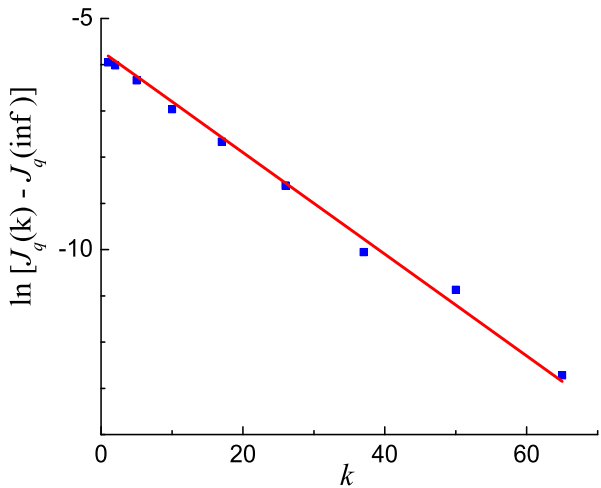} 
\vspace{-0.6cm}
\caption{(Color online) Dependence of $J_q(k)$ on $k$ obtained from simulation is plotted in squares and the line is a linear fit with slope equals -0.11. $q=0.2$, $\ell =12$ and $N=1000$.}
\label{fig:fit}
\end{figure}

Returning to Fig.~\ref{fig:one-J}, we note that significant deviations from the asymptotic value, $J_q(\infty )$, are found only for $k$'s within a small distance from the boundaries. At a casual glance, this range appears to depend on both $q$ and $\ell $. On closer examination of, say, the most prominent case here: $(q,\ell )=(0.2,12)$, we find that the decay of $J_q(k)$ into $J_q(\infty )$ fits an exponential quite well (Fig.~\ref{fig:fit}), i.e., 
$J_q(k)-J_q(\infty )\propto \exp \left( -k/\delta \right) $, with 
$\delta \approx 10$. Assuming this behavior persists in the other cases, we can study the $(q,\ell )$ dependence of this characteristic length and denote it by $\delta (q,\ell )$. We believe that the origin of this length scale can be traced to the presence of tails in the density profiles near the lattice boundaries: As the slow site approaches the entry point, these tails may seriously affect the injection process, and thus, the current as well. For the $\ell =1$ case, we observe that $\delta (q,\ell )\sim \xi (\beta _{eff})$, where $\xi $ is the characteristic length associated with the boundary layer of the \emph{density profile} in the ordinary TASEP. Specifically, with entry rate $\alpha =1$ and exit rate $\beta $, the system is in the H-phase and the profile decays exponentially into the bulk, as 
$\rho _x-\rho _{bulk}\propto \exp (-x/\xi )$ with \cite{Schutz}: 
\begin{equation}
\xi (\beta )=-\frac 1{\ln \left[ 4\beta (1-\beta )\right] }\,\,.
\label{H-xi}
\end{equation}
Here, the left half of our system is such a TASEP, except that we have an effective $\beta $: $\beta _{eff}=q/(1+q)$ \cite{LBS}. Using these
arguments on the three $q$'s shown, we estimate decay lengths of about $5$ ($q=0.4$), $3$ ($q=0.3$), and $2$ ($q=0.2$) lattice constants. Though the data on the differences $J_q(k)-J_q(\infty )$ are small and noisy, simulation results are consistent with 
$\delta (q,1)\sim \xi (\beta _{eff})$. However, for $\ell >1$, there is no analytic result for the boundary layers of the density profiles. Moreover, the data suggest that they are quite complex (e.g., in 
Figs.~\ref{fig:one-ribo-profile} and \ref{fig:one-profile-cf}). Thus, it is unclear how to quantify the picture for point particles to the general case of $\delta (q,\ell )$. At present, a complete understanding of both ``boundary layers'' -- in the density profiles and in $J_q(k)$ -- remains elusive. 

If we consider the edge effect as an ``interaction'' between the slow site and the lattice boundaries, the natural next step is to explore the interactions between \emph{two} slow sites. In order to avoid edge effects, we place the two slow sites sufficiently far away from the boundaries and vary their separation.

\subsection{Two slow sites}

\begin{figure}[tbp]
\hspace{-0.5cm} %
\includegraphics[height=3in,width=3.5in]{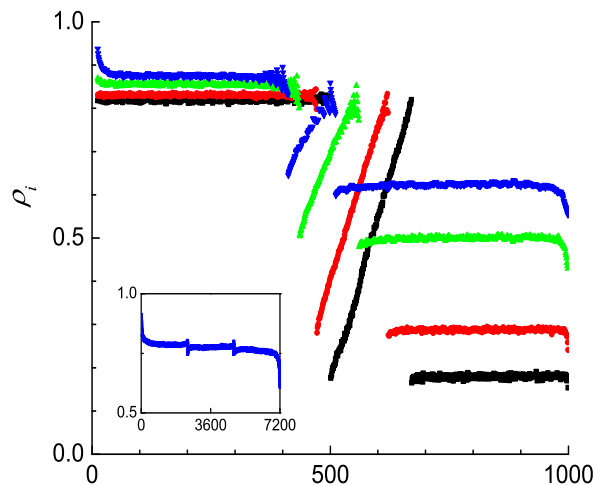} 
\vspace{-0.6cm}
\caption{(Color online) Coverage density profiles for two slow sites with $q=0.2$. $\ell =12$, $d=100$; $\ell =6$, $d=125$; $\ell =2$, $d=150$; and 
$\ell =1$, $d=170$ (Curves are plotted from left to right in the mid-subsection and top to bottom elsewhere). In all cases, $N=1000$. Inset: $q=0.6$, $\ell = 12$, $d= 2400$ and $N= 7200$.}
\label{fig:two-profile}
\end{figure}

As mentioned in the previous section, the currents in the $q_1\neq q_2$
cases are essentially controlled by the slower of the two rates and so, may be regarded as systems with a single slow site. These profiles can be interesting, but we choose to restrict our attention here to a study of the $q_1=q_2\equiv q$ case, in which the currents show a nontrivial dependence on $d$, the distance between the two slow sites. With two bottlenecks, the system consists of three sections: before the first blockage, in between the two, and after the second defect. Of course, for small $q$, the overall density in the first (last) section is expected to be high (low). In these cases, the effective entry and exit rates for the central section are also low, so that a wandering shock should be present. Hence, the \emph{average} profile should be linear for $\ell =1$ (and essentially so for larger $\ell $ \cite{LBSZia}) with a \emph{positive} slope. This behavior is understandable, since the section between the two defects is comparable to an ordinary TASEP with small $\alpha =\beta $. These expectations are generally confirmed by simulations with $q\lesssim 0.5$ and various $\ell $'s up to $12$. Fig.~\ref{fig:two-profile} shows typical coverage profiles, for a relatively small rate of $q=0.2$. The system appears to make a transition from this H/S/L phase to an M/M/M phase as $q$ increases. The center profiles become essentially flat, as illustrated in the inset (where $q=0.6$ and $\ell =12$). Details of this transition are being explored. 

\begin{figure}[tbp]
\hspace{-0.5cm} %
\includegraphics[height=3in,width=3.5in]{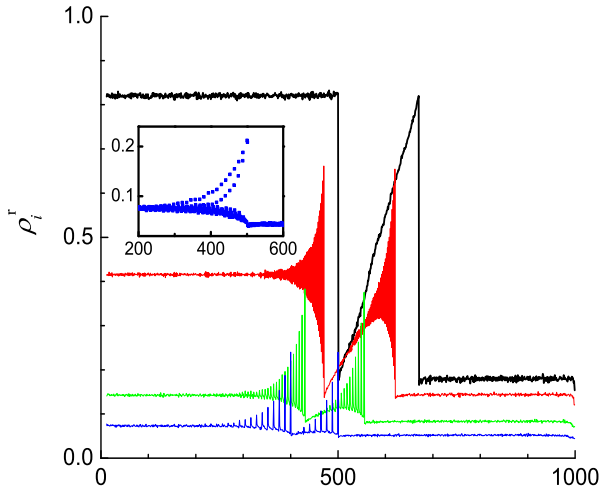} 
\vspace{-0.6cm}
\caption{(Color online) Ribosome density profiles with two slow sites of $q=0.2$. $\ell =12$, $d=100$; $\ell =6$, $d=125$; $\ell =2$, $d=150$; and 
$\ell =1$, $d=170$ (Curves are plotted from left to right in the mid-subsection and bottom to top elsewhere). Inset, $\ell =2$ and $d=1$. In all cases, $N=1000$.}
\label{fig:two-ribo-profile}
\end{figure}

More interesting are the finer features of the profiles in the small 
$q$ cases. As in the single defect system, the profiles exhibit period $\ell $ structures near the slow sites. To resolve these more clearly, we plot the \emph{reader} density profiles in 
Fig.~\ref{fig:two-ribo-profile}. In all cases that involve extended particles ($\ell >1$), the readers clearly pile up behind the slow sites. Apart from these ``jams,'' another feature emerges, namely, a sequence of \emph{depletion} zones, each of which precedes one
of the period $\ell $ peaks. For $\ell =2$, the differences between the
upper and the lower envelope are especially dramatic. More remarkably, when the blockages are separated by small $d$'s, two different, ``overlapping tails'' are created, as illustrated in the inset of
Fig.~\ref{fig:two-ribo-profile}, where $d=1$, $q=0.2$, and $\ell =12$.
Indeed, there are further interesting structures for 
$d\lesssim \ell $, which will be presented elsewhere. 

\begin{figure}[tbp]
\hspace{-0.5cm} \includegraphics[height=3in,width=3.5in]{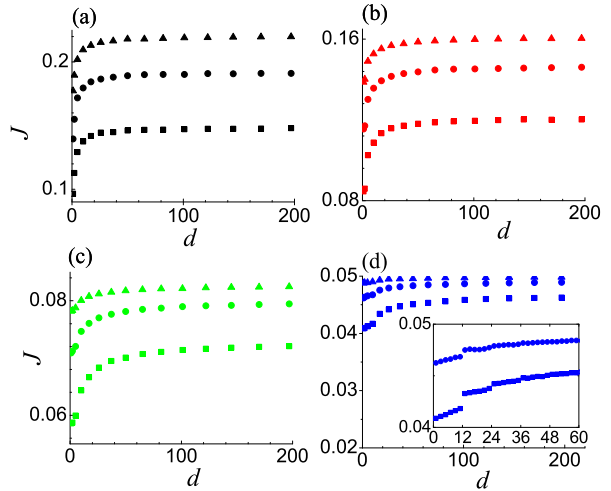} %
\vspace{-0.6cm}
\caption{(Color online) $J_q(d)$ as a function of the separation $d$ between the two slow sites for $q = 0.2$ (lower set of squares); $0.3$ (middle circles) and $0.4$ (upper triangles). (a) $\ell =1$; (b) $\ell =2$; (c) 
$\ell =6$; (d) $\ell =12 $. The inset in (d) is a magnified view of the $d\in \left[ 1,60\right] $ interval, to expose the period $\ell $ structures. In all cases, $N=1000$.}
\label{fig:two-J}
\end{figure}

\begin{figure}[tbh]
\hspace{-0.5cm} \includegraphics[height=3in,width=3.5in]{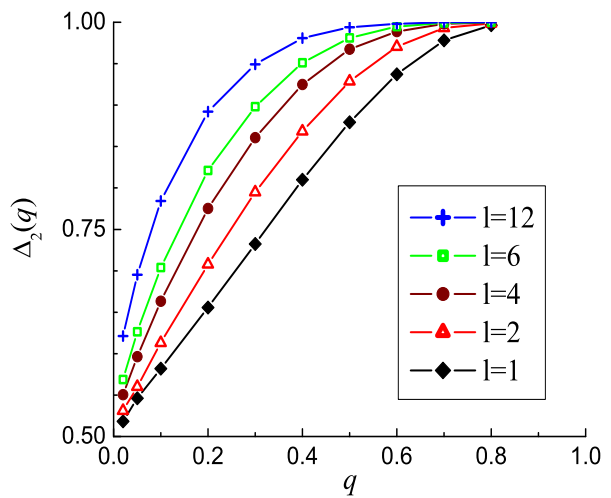} 
\vspace{-0.6cm}
\caption{(Color online) $\Delta _2(q)$ for $\ell =1,2,4,6$ and $12$.}
\label{fig:two-diff}
\end{figure}

Compared to these remarkable characteristics in the profiles, the behavior of the currents seems lackluster. In Fig.~\ref{fig:two-J}, we plot four sets of currents \cite{note3}, $J_q(d)$, associated with $\ell =1,2,6,$ and $12$. In all cases, we see that $J$ is considerably suppressed when $d$ is reduced. When the slow sites are very far apart, the current behaves as if there is only one slow site, consistent with expectations from mean-field theories. At the other extreme, when the two defect sites are nearest neighbors, the current reaches its minimum. Not surprisingly, period $\ell $ structures emerge as $d$ is varied, illustrated in the inset of Fig.~\ref{fig:two-J}(d), but become less prominent for $d\gtrsim 50$. These plots also reveal that, unlike the dependence on $k$ above, there are serious deviations from the 
$d\rightarrow \infty $ values when $d$ is decreased. To quantify this deviation, we define 
\begin{equation}
\Delta _2(q)=\frac{J_q(d=1)}{J_q(d\rightarrow \infty )}  \label{delta-J-2}
\end{equation}
and plot this quantity \emph{vs}.~$q$ in Fig.~\ref{fig:two-diff}. In contrast to $\Delta _1(q)$, we observe that $\Delta _2(q)$ exhibits a \emph{sizable} dependence on $q$, especially for small values of $q$. In the limit of $q\rightarrow 0$ the current decreases by a factor of 2! In the following section, we will see that this factor can be understood via a mean-field approach. 

To summarize our simulation results, two bottlenecks near each other have a dramatic effect on the current. We may regard this phenomenon as an ``interaction'' between the two slow sites, inducing far more ``resistance'' when they are close than when they are well separated.

\begin{figure}[tbp]
\hspace{-0.5cm} \includegraphics[height=3in,width=3.5in]{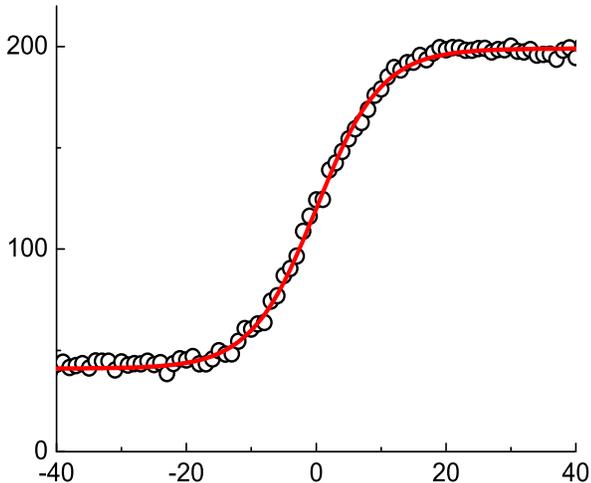} 
\vspace{-0.6cm}
\caption{(Color online) The open circles mark the average profile of a shock between sites $149$ and $349$ with $q=0.2$, compiled from of a very long run ($3\times 10^8$ MCS) with an $N =1000$, $\ell =1$ system. Details of how raw profiles are shifted (so that the shock is located at site $x=0$ shown here) will be published elsewhere \protect\cite{promises}. A simple fit using $A+B\tanh \left( x/10\right) $ is also shown: solid line. }
\label{fig:shock}
\end{figure}

Two additional comments are in order. First, we return to one of the
predictions of the mean-field theory, namely that a second slow site, spaced far apart from its partner, should have no further effect on the current. Our data indicate that the current for two slow sites, spaced far apart, is systematically \emph{lower} than the current for a single slow site, but only by a very small amount (less than $1\%$). Second, we can again attempt to identify a length scale which controls how $J_q(d)$ approaches $J_q(\infty )$, as $d$ increases. Since the central section of the system displays a shock, it is natural to ask whether the intrinsic width of the shock sets this length scale. According to \cite{Janowsky,JandL}, this width covers only a few lattice spacings in the \emph{periodic} TASEP with a single defect. Here, however, it appears that the shock is much broader. For example, the averaged profile of the shock for the case of $q=0.2$ with point particles is shown in Fig.~\ref{fig:shock}, as well as a simple fit using a 
$\tanh$ function \cite{note4} with width of about $10$. Intriguingly, this length appears to be comparable to the one appearing in 
Fig.~\ref{fig:two-J}(a). More work is needed to fully explore these issues.

\section{\label{sec:MF}Mean-Field Theoretic Approaches}

Mean-field theory is known to agree well with the exact results for a number
of macroscopic quantities in the steady state of the $\ell =1$ TASEP (see,
for example \cite{Schutz}). For extended particles, no exact solution is
available \cite{noExact} so that mean-field (and more sophisticated cluster)
approximations form the only route toward some understanding of the system's
behavior.However, there are \emph{many levels} of ``mean-field'' approximations \cite{MG,LBSZia,Lakatos,LBS}, corresponding to neglecting different types of correlations. For certain quantities (e.g., currents in large systems), predictions from the simplest level are very close to the simulation results. For others (e.g., some reader profiles), only the most sophisticated level performs adequately. In all cases, no level of mean-field theory can give a good fit to both the current and the profile. A thorough discussion of all of these schemes is quite involved and will be
provided elsewhere \cite{promises}. Here, we will restrict ourselves to the simplest method and compare its performance to the simulations.

All approaches start with the exact expressions for the current 
\begin{eqnarray}
J &=&\alpha \left\langle 1-n_\ell \right\rangle  \label{exactJ1} \\
&=&\gamma _i\left\langle r_i\left( 1-n_{i+\ell }\right) \right\rangle
\,\,;\quad i\in \left[ 1,N-\ell \right]  \label{exactJ2} \\
&=&\gamma _i\left\langle r_i\right\rangle \,\,;\quad \quad i\in \left[ N-\bar{\ell},N-1\right]  \label{exactJ3} \\
&=&\beta \left\langle r_N\right\rangle \,\,.  \label{exactJ4}
\end{eqnarray}
In the absence of the steady-state distribution, the most naive
approximation is to replace $\left\langle r_in_j\right\rangle $ by 
$\left\langle r_i\right\rangle \left\langle n_j\right\rangle $.
Unfortunately, the constraints due to particles with $\ell >1$ are so severe that this approximation is entirely inadequate when $j=i+\ell $. Even for the simple case of TASEP on a periodic ring, it leads to an erroneous expression for $J$ (except if $\ell =1$). Instead, the average (coverage) density at site $i+\ell $ is much larger than the (conditional) probability that it is actually covered \emph{given} that the reader is at site $i$. MacDonald and Gibbs (MG) proposed \cite{MG} a much better approximation: 
\[
\left\langle r_i\left( 1-n_{i+\ell }\right) \right\rangle \simeq \frac{\rho
_i^{\text{r}}(1-\rho _{i+\ell })}{1-\rho _{i+\ell }+\rho _{i+\ell }^{\text{r}%
}}=\frac{\rho _i^{\text{r}}\rho _{i+\ell }^{\text{h}}}{\rho _{i+\ell }^{%
\text{r}}+\rho _{i+\ell }^{\text{h}}}\,\,. 
\]
As discussed in Section \ref{sec:model}, the densities far from boundaries are uniform (e.g., 
$\rho _{i\rightarrow \infty }^{\text{r}}\rightarrow \rho_{bulk}/\ell $) and this fact provides a good description of the current-density relation 
\begin{equation}
J\left( \rho _{bulk}\right) =\frac{\rho _{bulk}\left( 1-\rho _{bulk}\right) 
}{\ell -\bar{\ell}\rho _{bulk}}  \label{J-rho}
\end{equation}
(for $\gamma =1$). Exploiting this relation and regarding our model as two or three TASEPs joined by slow sites, the simplest level of mean-field theories can be built. Ours is similar to, but simpler than, the approach in \cite{LBS} for the single defect case. The main difference lies in the matching condition, i.e., what approximate expression for the current across the slow site to use. After comparing the two approaches, we proceed to build the case for TASEP with two defects. 

\subsection{One slow site}

When a single slow site ($q<1$) is located at $k$, the system can be treated as two sublattices: $\left[ 1,k\right] $ and 
$\left[ k+1,N\right] $, referred to as the left and right sublattices, respectively. Associated quantities will appear with subscripts $L$ and $R$. The two sections are coupled through the slow site by having the same current in the steady state. Given this constraint, there are only two viable scenarios for the sublattices, out of the 3$\times $3 logically possible ones: H/L and M/M.

First, let us consider the H/L case which was one studied extensively in \cite{LBS}. The current for each sublattice can be written as: 
\begin{equation}
J_L=\frac{\beta _L(1-\beta _L)}{1+\beta _L\bar{\ell}}\quad J_R=\frac{\alpha
_R(1-\alpha _R)}{1+\alpha _R\bar{\ell}}\,,  \label{JLJR}
\end{equation}
where $\beta _L$ and $\alpha _R$ are the effective exit and entry rates, to be
determined later. By definition, the entire system reaches steady state when 
$J_L=J_R$, which yields $\beta _L=\alpha _R$. Of course, these are
intimately related to the (bulk) densities through $\rho _L=1-\beta _L$ and $\rho _R=\ell \alpha _R/\left( 1+\alpha _R\bar{\ell}\right) $, so that 
\[
\frac{\left( 1-\rho _L\right) }{1+\left( 1-\rho _L\right) \bar{\ell}}=\frac{\rho _R}\ell \,\,. 
\]
Another way to regard this relation is that both densities lead to the same current, which we denote by $J$ (a value to be determined, and equal to $J_L=J_R$). So, the high and low densities can be written as $\rho _{+}\left(J\right) $ and $\rho _{-}\left( J\right) $, respectively, being the two roots to Eqn.~(\ref{J-rho}). They will play a crucial role when we impose the matching condition, thereby fixing all quantities as a function of $q$.

The exact equation for ``matching'' is 
\begin{equation}
J=q\left\langle r_k\left( 1-n_{k+\ell }\right) \right\rangle \,\,,
\label{Jq_exact}
\end{equation}
in which $r_k$ and $n_{k+\ell }$ lie in $L$ and $R$, respectively. Now, the right can be expressed as, again exactly, 
$p\left( k|k+\ell \right)$, the probability for finding a ribosome at $k$, \emph{conditioned} on the presence of a hole at $k+\ell $.

Since we have ``broken'' the system into two separate TASEPs, a naive
approximation is to begin with 
\[
J_{NMF}=q\left\langle r_k\right\rangle \left\langle \left( 1-n_{k+\ell
}\right) \right\rangle 
\]
where the subscript stands for ``naive mean field.'' Regarding this as Eqn.~(\ref{exactJ1}) for the $R$ sublattice, we have 
$\alpha _R=q\rho _k^{\text{r}} $. Now, $\rho _k^{\text{r}}$ is in the $L$ sublattice, and must be related to $\rho _L$ in a mean-field approach. The most naive assumption is that $\rho _k^{\text{r}}$ is the same as its average in the bulk, i.e., $\rho_{bulk}^{\text{r}}$, which would be $\rho _L/\ell $ in this case. However, this turns out to underestimate $p\left( k|k+\ell \right) $ seriously. Indeed, the ``pile-up'' near a blockage (e.g., in Fig.~\ref{fig:one-profile-cf}) shows that $\rho _k^{\text{r}}$ is significantly higher than its bulk value as well as the densities on the $\bar{\ell}$ sites before. Thus, we propose that a better approximation would be to replace 
$\rho _k^{\text{r}}$ by $\rho _L$, and we write 
\begin{equation}
\alpha _R=q\rho _L\,\,.  \label{alphaR}
\end{equation}
Using $\rho _L=1-\beta _L$ and $\beta _L=\alpha _R$, so that $\alpha
_R=\beta _L=q/(1+q)$ and 
\[
\rho _L=1/(1+q)\,,\quad \rho _R=q\ell /(1+q\ell )\,, 
\]
we arrive at $J_{NMF}=q/\left[ (1+q)(1+q\ell )\right] $. The premise behind this line of arguments is that the system is in H/L, so that both $\alpha _R$ and $\beta _L$ should be less than $\hat{\chi}$ . Therefore, this expression for the current should be valid only if it is less than the maximal value ($\hat{\chi}^2$). In other words, the domain of its validity is limited to $q\leq 1/\sqrt{\ell }$. For higher $q$, this approach predicts that the system will be in an M/M phase, with maximal current. Note that such a phase cannot occur with a slow defect in the $\ell =1$ case, where M/M can be accessed only with 
$q>1$. In an earlier study \cite{LBS}, the parameters chosen ($q=0.2$ and $\ell =12$) also precluded the presence of this phase,
although we believe (see below) that this phase cannot be present if the blockage is in the center ($k=N/2$) or deep in the bulk. We summarize this ``naive mean field''
by 
\begin{equation}
J_{NMF}=\left\{ 
\begin{array}{cc}
q/\left[ (1+q)(1+q\ell )\right] & \text{ for }q\leq 1/\sqrt{\ell } \\ 
\hat{\chi}^2 & \text{ for }q\geq 1/\sqrt{\ell }
\end{array}
\right. \,.  \label{J_NMF}
\end{equation}

An alternative approximation for eqn. (\ref{Jq_exact}) was proposed earlier \cite{LBS}: 
\begin{equation}
J\cong q_{eff}\left( \frac{\rho _L}\ell \right) \left( \frac{1-\rho _R}{1-\rho _R\bar{\ell}/\ell }\right) \,\,.  \label{SKL-J}
\end{equation}
The last two factors can be recognized as $\left\langle r_k\right\rangle $ and the MG approximation for the effective hole density \cite{MG}. The first factor is a little more subtle \cite{LBS}: Considering that the transit time for a single particle through the slow site (in the absence of steric hindrance) is $q^{-1}+\bar{\ell}$, $q_{eff}$ is defined as the average rate to move just one step in this process: 
\[
q_{eff}\equiv \frac{q\ell }{1+q\bar{\ell}}\,\,. 
\]
The end result for the current is the solution to the algebraic equation 
\[
J=q_{eff}\frac{\rho _{+}\left( J\right) \left[ 1-\rho _{-}\left( J\right)
\right] }{\ell -\rho _{-}\left( J\right) \bar{\ell}}\,\,. 
\]
Here, we give an explicit form (which displays the $\ell =1$ limit well) 
\begin{equation}
J_{SKL}=\frac Q{\left( 1-Q+\sqrt{1-2Q}\right) \bar{\ell}}  \label{J_SKL}
\end{equation}
where 
\[
Q\equiv \frac{2q\bar{\ell}\left( 1+q\bar{\ell}\right) }{\left( 1+q+2q\bar{\ell}\right) ^2}\,\,. 
\]
Note that, for any $q<1$, this approach predicts that the current is less than the maximal value of $\hat{\chi}^2$ and so, the system is always in the H/L phase.
\begin{figure}[tbp]
\hspace{-0.5cm} \includegraphics[height=3in,width=3.5in]{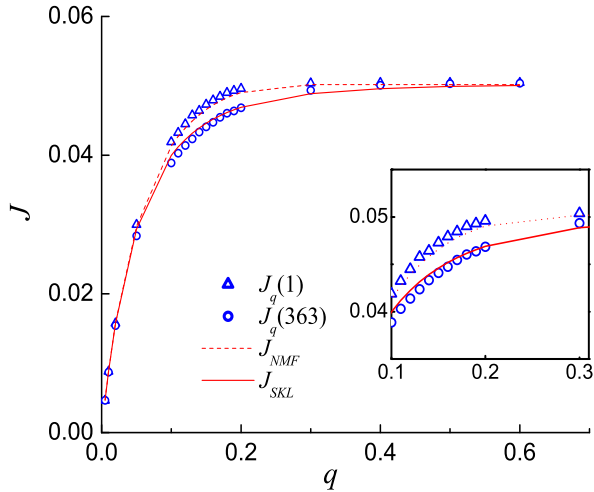} 
\vspace{-0.6cm}
\caption{(Color online) Comparisons of the current, $J$, as a function of $q$. The legend labels the two sets of simulation data (slow site at $k=1$ and $363$) and predictions from two mean-field approximations. }
\label{fig:J(q)}
\end{figure}

The results of both mean-field predictions for $\,J$ as a function of $q$ are shown in Fig.~\ref{fig:J(q)}, along with two sets of data: 
$J_q\left(1\right) $ and $J_q\left( 363\right) $. As expected, 
$J_{SKL}$ was purpose-built for two infinite TASEP's connected by a slow site and provides a better fit to the data with the blockage deep in the bulk ($k=363$ in $N=1000$). On the other hand, it is understandable that, e.g., for $k=1$, $\alpha _R$ must
be very close to $q$. Thus, we may expect that the system will have maximal current for $q\gtrsim 1/\sqrt{\ell }$. This behavior is confirmed by the data, as illustrated in the figure for $\ell =12$. Since $J_{NMF}\left( q\right) $ has the property that it saturates at $\hat{\chi}^2$ for $q>\hat{\chi}$, it provides a better fit for $J_q\left( 1\right) $. Of course, we recognize that, as mean-field
theories, neither (\ref{alphaR}) nor (\ref{SKL-J}) are the first step in a systematic expansion, so that they may better be thought of as
``semi-phenomenological''.

More importantly, there is a more serious, inherent limitation in this level of mean-field theory. It cannot account for the full $k$-dependence in $J_q\left( k\right) $, since it deals only with infinite systems. One possibility to incorporate finite-size effects (of the left sublattice) in this kind of theory is to exploit the MG expression \cite{MG} 
\[
J_{MG}=\frac{\rho _i^{\text{r}}\rho _{i+\ell }^{\text{h}}}{\rho _{i+\ell }^{%
\text{r}}+\rho _{i+\ell }^{\text{h}}} 
\]
for sites near the boundaries. Using it in a recursion relation for finding both the density profile and the current, we will arrive at a $k$-dependent expression, $J_{MG}\left( \alpha =1,\beta _L;k\right)$, 
which replaces $J_L$ in the first expression of Eqn.~(\ref{JLJR}). However, the high density fixed point of this recursion relation is unstable and very careful numerical analysis will be necessary. Preliminary work on this approach is promising and will be reported elsewhere \cite{promises}.

We end this subsection by noting that the effects of a single defect in
TASEP have also been investigated in \cite{HadN}. Unlike our focus here - the dependence of $J$ on the \emph{location} of the slow site, they are concerned with a ``multicritical system,'' i.e., $\alpha =\beta =1/2$ for the $\ell =1$ case. Putting the defect at the center of the lattice, they explored density profiles in detail, finding power law tails on both sides of the defect with $q$-dependent exponents. By contrast, our choice of $\alpha =\beta =1$ places us far from the multicritical point. We have no reason to expect similar power laws.

\subsection{Two slow sites}

The most general TASEP with just two slow sites can be quite involved, since the parameter space is four-dimensional: $\left\{ q_1,q_2,k_1,k_2\right\} $. To carry out a manageable investigation, we let both sites be deep in the bulk, so that only the distance between them, $d\equiv (k_2-k_1)$, plays a significant role. Further, as pointed out above, the central section resembles an ordinary TASEP with $\alpha $ and $\beta $ controlled by $q_1$ and $q_2$, respectively. Therefore, it is the smaller (slower) of the two rates which limits that current, which in turn dictates the current through the whole system. Thus, we will focus only on the $q_1=q_2=q$ case. Our parameter space will then
\emph{resemble} the single slow site case.

Following the single defect case, the simplest levels of mean-field theory treat our system as three subsections with obvious labels: $L,C,$ and $R$. From our discussion, only two (out of the many logical possibilities) combinations of phases, H/S/L and M/M/M, are expected to be viable. In addition to Eqn.~(\ref{JLJR}), we have 
\begin{equation}
J_C=\frac{\alpha _C(1-\alpha _C)}{1+\alpha _C\bar{\ell}}\,.  \label{J_C}
\end{equation}
Since the defect rates are identical, we fully expect that, for such a mean-field theory, $\beta _C=\alpha _C$. Now, matching the currents of the subsections, we immediately arrive at $J_L=J_C=J_R$ and so, 
$\beta _L=\alpha_C=\beta _C=\alpha _R$. From here, the ``naive'' mean-field approach for H/S/L proceeds identically to the above. The argument relies on the presence of a shock in the central section, so that there is a low (high) density region near site $k_1+1$ ($k_2$) and we can impose the same discontinuity in the densities across both defects, i.e., $\rho _{+}\left( J\right) $ before and 
$\rho _{-}\left( J\right) $ after. Thus, we again arrive at 
$J_{NMF}\left( q\right) $, given explicitly in Eqn.~(\ref{J_NMF}).
The same argument can be applied to the next level of a mean-field
approximation, which predicts $J_{SKL}\left( q\right) $, as in 
Eqn.~(\ref{J_SKL}). The major difference between the two approaches, as in the single slow site case, is the absence of the M/M/M phase in the latter. Meanwhile, their limitations are similar: The $d$ dependence in $J_q\left( d\right) $ cannot be accommodated without serious modifications.

Nevertheless, the spirit of these approximations can be exploited to provide $J_q\left( 1\right) $ in the $q\rightarrow 0$ limit. Since the central section consists of just one site, there can be no shock. Instead, the remnant of the shock is reflected in the average density there. It is more convenient to regard the system as two infinite TASEP's, with nontrivial matching across a ``doubly - slow site.'' Of course, we cannot expect to find any of the fascinating profile details (e.g., inset of Fig.~\ref{fig:two-ribo-profile}), but we should be able to obtain the ``coarser'' information, such as the currents. The goal is to understand the behavior of $\Delta_2(q\rightarrow 0)$ (cf.~
Fig.~\ref{fig:two-diff}) in, say, the first two non-vanishing orders in 
$q$.

Now, for $q\ll 1$, we are naturally in the H/L phase and the crudest
approximation should suffice for the lowest order in the current. So, we let the bulk densities be at their extremes (i.e., one and zero) and simply consider the time it takes for a particle to move through the blockage, from the moment its predecessor is ``released.'' The current is just the inverse of this quantity, i.e., 
\begin{equation}
\left[ \frac 2q+\ell -2\right] ^{-1}\rightarrow \frac q2\left[ 1-\frac
q2(\ell -2)+...\right] \,\,,  \label{q/2}
\end{equation}
where we have included $O\left( q^2\right) $ terms for computing the next order. But, at this order, we should also take into account that,
occasionally, the density before/after the blockage deviates from unity/zero by virtue of the right hand side of Eqn.~(\ref{J-rho}) being non-zero. Thus, these densities are 
\begin{eqnarray*}
\rho _L &\rightarrow &1-J=1-q/2+... \\
\rho _R &\rightarrow &J\ell \approx q\ell /2+...
\end{eqnarray*}
and further suppress the current at the next-to-lowest order
through the factor 
\begin{equation}
\rho _L\left( 1-\rho _R\right) \rightarrow 1-\frac q2(\ell +1)+...\,\,.
\label{dev-2}
\end{equation}
Combining these factors, we arrive at 
\[
J_{q\rightarrow 0}\left( d=1\right) \rightarrow \frac q2\left[ 1-q %
\left(\ell - \frac 12 \right) +...\right] 
\]
If we use exactly the same arguments for the $q\rightarrow 0$ limit current in the case of one slow site, we find, instead of (\ref{q/2}), 
\begin{equation}
\left[ \frac 1q+\ell -1\right] ^{-1}\rightarrow q\left[ 1-q(\ell
-1)+...\right] \,\,,  \label{q/1}
\end{equation}
and, instead of (\ref{dev-2}), 
\[
\rho _L\left( 1-\rho _R\right) \rightarrow 1-q(\ell +1)+...\,\,. 
\]
Finally, since $J_q\left( d\rightarrow \infty \right) $ is the same as the
single-blockage current, we write 
\begin{equation}
J_{q\rightarrow 0}\left( d\rightarrow \infty \right) \rightarrow q\left[
1-2q\ell +...\right]  \label{dev-1}
\end{equation}
so that 
\[
\Delta _2(q\rightarrow 0)\rightarrow \frac 12+\frac q2(\ell +\frac 12)+...\,\,. 
\]

It is remarkable how well this crude approximation agrees with the data in
Fig.~\ref{fig:two-diff}. There is no doubt that all curves extrapolate to the $\ell $-independent value of $1/2$ at $q=0$. As for the slope at the origin,
we can obtain a good estimate from the lowest $q$ data points, using $\left[
\Delta _2(q=0.02)-0.5\right] /0.02$. The values obtained from simulations for $\ell =1,2,4,6,$ and $12$ are 
$0.92,$ $1.55,2.53,3.44,$ and $6.06$, respectively.

We are aware that the expansion (\ref{dev-1}) differs from the small 
$q$ limit of $J_{NMF}$. Unfortunately, it is difficult to implement the same scheme for $J_{NMF}$ here, since we must start from the exact \emph{pair} of equations: 
\begin{equation}
J=q\left\langle r_k\left( 1-n_{k+\ell }\right) \right\rangle =q\left\langle
r_{k+1}\left( 1-n_{k+\ell +1}\right) \right\rangle \,\,.  \label{J2q_exact}
\end{equation}
Various attempts at approximating $\rho _k^{\text{r}}$ or 
$\rho _{k+1}^{\text{r}}$ led to poorer results. Alternatively, we could exploit the argument in SKL \cite{LBS} and consider the average time to traverse both slow sites, $2/q+\left( \ell -2\right) $. This gives us a new effective $q$: 
\[
\tilde{q}_{eff}\equiv \frac{q\ell }{2+q\left( \ell -2\right) } 
\]
which can be inserted into Eqn.~(\ref{SKL-J}). The result is $\Delta
_2(q\rightarrow 0)\rightarrow \frac 12+\frac q4\left( \ell +2\right) +...$, the $O\left( q\right) $ term of which differs from the data by about a factor of 2. Clearly, mean-field approaches are far from ideal for finding quantitative predictions of $J_q\left( d\right) $. On the other hand, either $J_{NMF}\left( q\right) $ or $J_{SKL}\left( q\right) $ provide tolerable results when the blockages are from from each other or the boundaries. Such variations in the quality of mean-field theories point to the importance of correlations. Considerable efforts appear to be necessary for a comprehensive, yet relatively simple, theory.

Let us end this section with another method which could possibly improve the theoretical predictions \cite{TCidea}, especially for the first few values of $k$. The idea is to find exact results, by solving the full master equation, for TASEPs (with extended objects) on very small lattices and then to match these to an infinite system (the $R$ sublattice). One expectation is that, as in the $\ell =1$ case, the finite-size current is larger than its counterpart for the infinite TASEP at the same $(\alpha ,\beta )$. This approach may eventually provide the essential argument to understand the increase in $J_q\left( k\right) $ as $k$ becomes smaller. Similarly, this idea can be applied to the case with two slow sites when $d$ is $O\left( 1\right) $. Work is in progress to explore these avenues.

\section{\label{sec:SUM}Summary and conclusions\protect\newline}

In this study, we consider an inhomogeneous TASEP with open boundaries and populated with particles of finite extent, $\ell $. The hopping rates are uniform (set at unity) \emph{except} for one or two sites (``defect bonds''), where the rates, $q$, are different (faster or slower). We are interested in the effects of these local defects on the density profiles and the currents through the system. Simulations with various $\ell \leq 12$ show that fast sites have no effect on the current, but induce discontinuities in the density profiles. In contrast, slow sites generate a significant reduction of the current as well as nontrivial structures in the profiles, e.g., long tails behind the blockage, with period $\ell $. These findings are entirely consistent with similar studies in the past, most of which were restricted to $\ell =1$ \cite{Kolo,HadN}. If the inhomogeneities are
deep in the bulk and far from each other, the current depends only on $q$ and can be understood through simple mean-field considerations. Through the current-density relationship (for an infinite homogeneous TASEP), the overall densities in each of the defect-free sections can be also predicted, so that the various ``phases'' of these subsystems can be understood. The distinguishing feature in our study is how the\emph{\ location} of the defects affects the behavior of the system. For the case of one slow site, the current is slightly but measurably enhanced when the defect approaches the boundary. On the
other hand, a drastic reduction of the current is observed when two slow sites are brought closer to each other. It is tempting to interpret these effects as ``interactions'' between the defects
and to seek a formulation that can describe them quantitatively. 

At present, neither the enhancement nor the suppression of the currents can be understood in terms of simple mean-field theories. The essential
limitation is that they are based on matching homogeneous TASEPs of \emph{infinite} length. More sophisticated versions, relying on recursion relations for the particle density at each site, may be exploited to deal with the finite subsections and provide some promise for a better understanding of such effects. For very small subsections, such as $k,d\lesssim 5$, it may be possible to find exact solutions (even for $\ell >1$) that can be used to match the mean-field descriptions for the macroscopic subsection(s). Work is in progress to investigate these approaches systematically.  

Beyond one or two blockages, we should study systems with multiple
slow sites, as our eventual goal is to understand the properties of fully inhomogeneous TASEPs. In particular, though restricted to just one or two slow sites, our findings - that the relative locations of blockages are important - will have implications for translation. For example, they are directly applicable to ``designer genes'', which consist of many repeats of the same codon, except at one or two locations. Using the abundance of associated aa-tRNAs as a control for the elongation rate across any particular codon, we can test our results directly on such genes. Thus, it will be interesting to see the physical manifestation of, e.g., enhancement and suppression of production rates of such artificial proteins, depending
on the placement of the slow codon(s). Similarly, reproducing the intriguing ribosome density profiles will be revealing. More important than ``designer genes,'' we should consider the implications for real genes. Our results should provide, at the least, some simple qualitative insights. We can obviously maximize the production rate of a particular protein associated with a certain real gene by systematically replacing all slow codons with synonymous, faster ones. However, for most genes this operation will require a large number of replacements. Instead, with our findings in mind, we can achieve considerable increases in the production rates by making only a
few substitutions, namely, by replacing the slowest codons, or a cluster of nearby slow codons. The ratio of current enhancement to the number of codon replacements may be used to quantify how ``optimal'' a certain set of substitutions is. This idea can be applied to finding optimal means to \emph{suppress} prodcution rates as well. Simulation work with TASEPs associated with real genes is in progress and we hope to demonstrate that these concepts are viable. 
\begin{acknowledgments}
We have benefited from discussions with T. Chou, M. Evans, M. Ha, R. Kulkarni, P. Kulkarni, M. den Nijs, S.-C. Park, L.B. Shaw, and B. Winkel. We are especially grateful to M. Ha and M. den Nijs for providing their unpublished data. This work is supported in part by the NSF through DMR-0414122, DMR-0705152, and DGE-0504196. JJD also acknowledges the generous support from the Virginia Tech Graduate School.
\end{acknowledgments}


\begin{thebibliography}{99}
\bibitem{note1}  In much of the physics literature, the term ``bond'' is used instead of ``site'', since hopping is associated with a particle jump from site $i$ to site $i+1$. However, in translation, a ribosome ``at site $i$'' can move to the next site only when the tRNA associated with site $i$ arrives. Therefore, it is natural to associate the jump rate with a site, and so we will use terms like ``slow site'' and ``slow bond'' interchangeably. With protein synthesis in mind, we also use the phrase ``slow codon.''
\bibitem{note4}  Since the shock diffuses throughout the region between the slow sites, a relatively nontrivial method must be used to compile
averages. Details will be published elsewhere \cite{promises}
\end{thebibliography}
\end{document}